\documentstyle[twoside,fleqn,espcrc2,epsf]{article}

\newcommand{\AmS}{{\protect\the\textfont2
  A\kern-.1667em\lower.5ex\hbox{M}\kern-.125emS}}

\newcommand{\bee}{\begin{equation}}
\newcommand{\ee}{\end{equation}}
\newcommand{\beea}{\begin{eqnarray}}
\newcommand{\eea}{\end{eqnarray}}

\title{Understanding chiral symmetry breaking with the overlap action}

\author{Thomas DeGrand \\[2mm]
        Department of Physics, 
        University of Colorado\\ Boulder, CO 80309-390, USA}

\begin{document}

\begin{abstract}
A chiral fermion action allows one to do very clean studies of chiral symmetry
breaking in QCD. I will briefly describe how to compute with the overlap action
(relatively) cheaply, and then turn to physics:
Low modes of the Dirac operator show a ``lumped'' chiral density which
peaks at the locations of instantons and anti-instantons. These modes
dominate  correlation functions at small quark mass in many channels.
The picture qualitatively (and in some cases quantitatively) resembles
an instanton liquid model.
\end{abstract}

\maketitle
Instantons are a
 plausible candidate for the source of chiral symmetry breaking.
 Would-be fermion zero modes sitting on individual instantons
mix into a band, and the density of low modes is connected
to the chiral condensate via the Banks-Casher relation.
An elaborate phenomenology built on the interactions of fermions
with instantons can account for many of the low energy
properties of QCD \cite{ref:reviews}. But is this picture true?

In principle, lattice simulations can address this issue.
 However, the usual lattice fermion actions themselves distort chiral symmetry.
This can cloud the lattice results, since what is observed might just be due
to the bad behavior of the action.
Lattice actions
which implement an exact chiral symmetry without doubling
(namely, the overlap action \cite{ref:neuberfer})
allow one to study these questions in a theoretically clean context.
This note describes recent work by Anna Hasenfratz and me, studying
chiral symmetry breaking using an overlap action.

A generic overlap operator is
\bee D(0) = x_0(1+ {z \over{\sqrt{z^\dagger z}}} )
\label{eq:gw}
\ee
where $z = d(-x_0)/x_0$ and $d(m)=d+m$ is some massive nonchiral
 Dirac operator for mass $m$.
My\cite{TOM1} 
 choice of ``kernel'' $d$ is an action with nearest and next-nearest
 neighbors and fat links as a gauge connection, designed to ``look like''
an overlap action.
I find eigenmodes of the squared  Dirac operator $D^\dagger D$ 
using a conjugate gradient 
routine\cite{ref:eigen}. This algorithm takes a set of
trial vectors and iteratively improves them (with many multiplications
of trial vectors by $D^\dagger D$).
 Any way it is implemented, an overlap action is much
more expensive to use than a non-overlap action.
 The crucial trick I use to speed  up the calculation is to begin the 
computation of eigenmodes of $D^\dagger D$
 with eigenmodes of $d(0)^\dagger d(0)$. If these modes
are close to eigenmodes of $D^\dagger D$,
 fewer iterations are needed.
This can give a gain of up to a factor of 20 in time needed
to find eigenvectors,  compared to the cost of finding eigenmodes of
the overlap with $d$ given by the Wilson action.
In principle this would work for any action, but in practice I could
not discover a good approximation to begin the calculation of eigenmodes
of the overlap with Wilson or clover kernels.

All of the studies reported here were done in quenched approximation at
$\beta=5.9$ (lattice spacing 0.11-0.13 fm), on $12^4$ and $12^3\times 24$ 
lattices.

\begin{figure}[h!tb]
\begin{center}
\leavevmode
\epsfxsize=70mm
\epsfbox{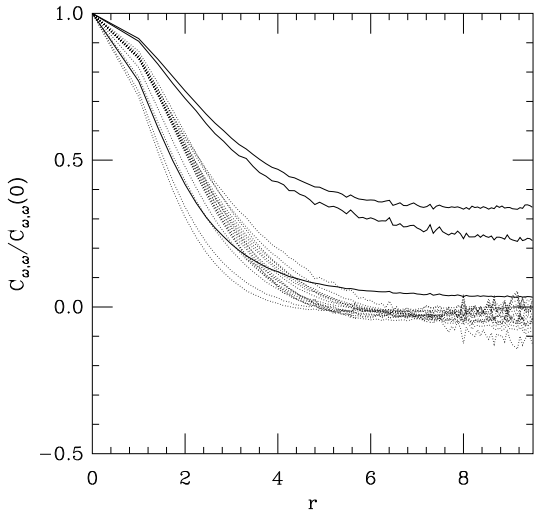}
\end{center}
\vspace{-28pt}
\caption{Chirality autocorrelator, normalized at the origin, and showing
chiral zero modes (the upper lines) and nonchiral modes.}
\label{fig:t1}
\end{figure}

We \cite{TOM2} collected a set of eigenmodes, extracted the
local chiral density $\omega(x)=\langle \psi(x)|\gamma_5|\psi(x)\rangle$
for each mode, and measured autocorrelation functions
 $C_{\omega\omega}(r)=\langle \omega(r)\omega(0) \rangle$ 
and correlators of the chiral density with the topological
charge density $Q(r)$ (as measured from a pure gauge observable),
 $C_{\omega Q}(r)=\langle \omega(r)Q(0) \rangle$. All modes showed
a strong peaking in both correlators at small $r$.
 Chiral zero modes ``sit'' on one sign
of bumps of topological charge, while nonchiral fermion modes are localized
on both signs of topological charge. As the fermion eigenvalue rises, these
correlations slowly die away, but they persisted out to eigenvalues
of 500 MeV or so. An example of a chirality autocorrelator is shown
 in Fig. \ref{fig:t1}.

We could  extract a density profile from the shape of the bumps.
Interpreting this profile a being due to a mixture of fermion zero modes
sitting on instantons, we infer a typical instanton radius of 
$\langle \rho \rangle \simeq 0.3$ fm, a familiar number from instanton
phenomenology.

But are these few modes (typically 10-20 out of $12^5$) important?
That is a qualitative question, but it is relevant to phenomenology.
To test this, we constructed quark propagators ``exactly,'' and 
propagators truncated to include only a few low eigenmodes. We then
looked at ordinary $\vec k=0$ propagators (as would be used in spectroscopy).
I also computed \cite{TOM3} point-to-point correlators, 
\bee
\Pi_i(x) = {\rm Tr}\langle J_i^a(x)J_i^a(0)\rangle .
\ee
In the latter case it is customary to divide out by the free field
correlator  $\Pi_i^0(x)$ and measure
\bee
R_i(x) = \Pi_i(x)/\Pi_i^0(x).
\label{eq:ratio}
\ee

When the quark mass was small enough (pseudoscalar/vector mass ratio 0.5
or less), a few modes were sufficient to saturate the pseudoscalar and scalar
correlators. This was not the case for the vector or axial channels.
Baryon channels are noisy, but low modes seemed to make a substantial
 contribution there, too. What we saw was basically consistent with instanton
liquid phenomenology: 
channels with spin-0 $q \bar q$ pairs or diquarks are supposed to couple
strongly to instantons, as the pair propagates by hopping from instanton
to instanton. Presumably this means that quark eigenmodes which are most 
sensitive to topology dominate the correlator.
In channels with spin-1 diquarks or $q \bar q$ pairs, the fermions cannot
couple to the same instanton, suggesting that low modes do not saturate
their correlators.

\begin{figure}[h!tb]
\begin{center}
\leavevmode
\epsfxsize=70mm
\epsfbox{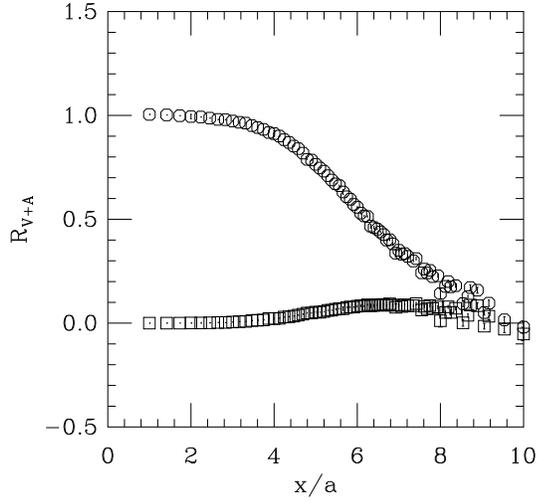}
\end{center}
\vspace{-28pt}
\caption{Comparison of $R_{V+A}$ from ``exact''
propagators  (octagons) and from propagators truncated to the lowest 10 modes
of $D^\dagger D$ (squares)
 at $m_{PS}/m_V=0.5$, on $12^4$ lattices at $\beta=5.9$.}
\label{fig:np}
\end{figure}

\begin{figure}[h!tb]
\begin{center}
\leavevmode
\epsfxsize=70mm
\epsfbox{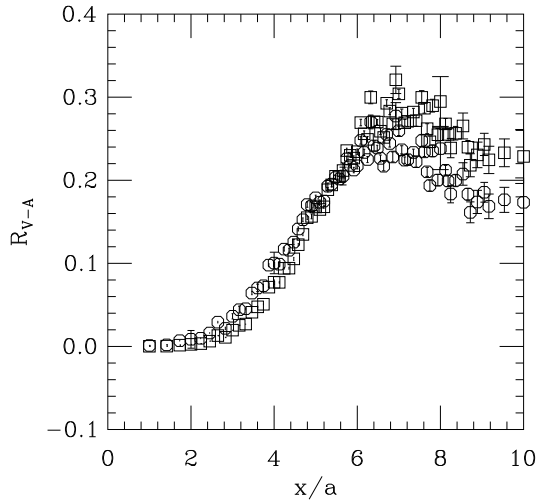}
\end{center}
\vspace{-28pt}
\caption{Comparison of $R_{V-A}$ as in Fig. \protect\ref{fig:np}.}
\label{fig:nn}
\end{figure}

As one example among many, consider the
 the correlator ratios $R_{V+A}$ and $R_{V-A}$.
In the sum rule/operator product expansion approach
$R_{V + A}$ is dominated by perturbative physics and is expected
to take a value very close to unity,
while $R_{V - A}$ is zero at small $x$ and receives only nonperturbative
contributions which are relevant to chiral symmetry breaking.
A comparison of full and low mode truncated correlators
 (at $m_{PS}/m_V \simeq 0.5$) is shown in Figs.~\ref{fig:np} and \ref{fig:nn}.
Where phenomenology expects instanton-sensitive modes, there they dominate.
An extrapolation of this lattice data to zero quark mass is in reasonably good
agreement with an instanton model \cite{ref:ssa1}
 and with tau-decay data.

\begin{figure}[h!tb]
\begin{center}
\leavevmode
\epsfxsize=70mm
\epsfbox{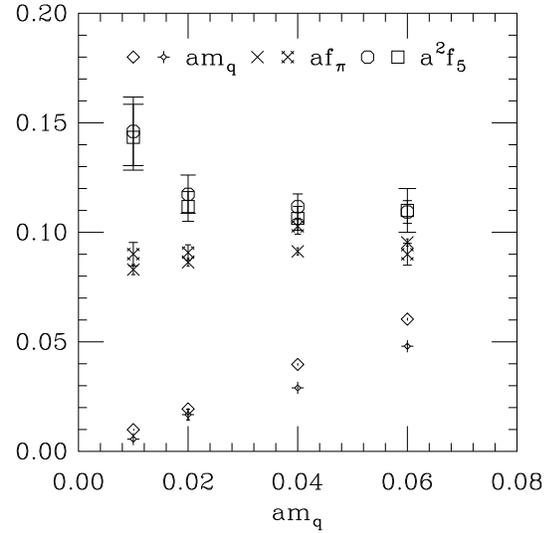}
\end{center}
\vspace{-28pt}
\caption{Comparison of pseudoscalar observables from ``exact''
propagators (the left of the two symbols)
and from propagators truncated to the lowest 20 modes
of $D^\dagger D$ (the right hand symbols), on $\beta=5.9$ $12^3\times 24$
configurations.}
\label{fig:t3}
\end{figure}

Finally, in Fig. \ref{fig:t3} I show a comparison of several pionic
matrix elements computed with complete quark propagators and from quark 
propagators truncated to the lowest 20 eigenmodes of $D$. Shown are
the PCAC quark mass (from the divergence of the axial vector current),
$f_\pi$, and $f_5=\langle 0 | \bar\psi \gamma_5 \psi |\pi\rangle$.
Low modes make a large contribution to these observables.

To conclude: Low modes of fermions show structure which is strongly
reminiscent of the expectations of instanton liquid models.
Low modes show a peaked chirality density correlated with gauge field topology.
These low modes contribute strongly in the pseudoscalar  channel
and other places where chiral symmetry is important,
when the quark mass is small. It seems that instantons are
connected, after all, with chiral symmetry breaking.

This work was supported by the US Department of Energy.


\begin{thebibliography}{9}
\bibitem{ref:reviews}
T.~Schafer and E.~V.~Shuryak,
Rev.\ Mod.\ Phys.\  {\bf 70}, 323 (1998)
[hep-ph/9610451];
%
D.~Diakanov, Lectures at the Enrico Fermi School in Physics,
Varenna, 1995 [hep-ph/9602375]
%
\bibitem{ref:neuberfer}
H.~Neuberger,
Phys.\ Lett.\  {\bf B417}, 141 (1998)
[hep-lat/9707022],
Phys.\ Rev.\ Lett.\  {\bf 81}, 4060 (1998)
[hep-lat/9806025].
%
\bibitem{TOM1}
T.~DeGrand  [MILC collaboration],
Phys.\ Rev.\ D {\bf 63} (2001) 034503
[hep-lat/0007046].
%
\bibitem{ref:eigen}
See B. Bunk, et al, unpublished DESY report 
(1994); T. Kalkreuter and H. Simma, Comp. Phys. Comm. 93, 33 (1996).
%
%
\bibitem{TOM2}
T.~DeGrand and A.~Hasenfratz,
Phys.\ Rev.\ D {\bf 64} (2001) 034512
[hep-lat/0012021].
\bibitem{TOM3}
T.~DeGrand,
hep-lat/0106001.
%
%
\bibitem{ref:ssa1}
T.~Schafer and E.~V.~Shuryak,
Phys.\ Rev.\ Lett.\  {\bf 86}, 3973 (2001)
[hep-ph/0010116].
%
\end{thebibliography}
\end{document}